\documentclass[english,letterpaper,aps,prl,twocolumn,superscriptaddress,groupedaddress,showpacs]{revtex4}
\usepackage{graphicx}
\usepackage{amsmath}
\usepackage{amssymb}
\usepackage{url}
\usepackage{float}
\usepackage{color}
\usepackage{natbib}\begin{document}

\title{Valley Order and Loop Currents in Graphene on Hexagonal Boron Nitride}

\author{Bruno Uchoa}
\affiliation{Department of Physics and Astronomy, University of Oklahoma,
Norman, OK 73069, USA}

\author{Valeri N. Kotov}
\affiliation{Department of Physics, University of Vermont, Burlington,
Vermont 05405, USA}

\author{M. Kindermann}
\affiliation{School of Physics, Georgia Institute of Technology, Atlanta,
Georgia 30332, USA}

\date{\today}
\begin{abstract}
In this letter, we examine the role of Coulomb interactions in the
emergence of macroscopically ordered states in graphene supported
on hexagonal boron nitride substrates. Due to incommensuration effects
with the substrate and interactions, graphene can develop gapped low
energy modes that spatially conform into a triangular superlattice
of quantum rings. In the presence of these modes, we show that Coulomb
interactions lead to spontaneous formation of chiral loop currents
in bulk and to macroscopic \emph{spin-valley} order at zero temperature.
We show that this exotic state breaks time reversal symmetry and can
be detected with interferometry and polar Kerr measurements. 
\end{abstract}


\maketitle
\emph{Introduction. }In spite of the presence of quasiparticles with
Dirac cone spectrum \cite{Antonio}, the emergence of topological
order in graphene is hindered by the fermionic doubling problem, where
electrons have a four-fold degeneracy in valleys and spins \cite{Haldane}.
Due to the vanishingly small density of states (DOS) at the Dirac
points, many-body instabilities in general are quantum critical and
require strong coupling regimes \cite{Kotov}. We argue that one promising
possibility to generate many body states that lift the fermionic degeneracy
and break time reversal symmetry (TRS) is to use substrates to reconstruct
the DOS of graphene near the Dirac points into nearly flat bands. 

In incommensurate two-layer crystals with honeycomb structure, the
Dirac points are protected by a combination of parity and TRS \cite{Fu}.
On top of hexagonal boron nitride (BN), where inversion symmetry is
broken, graphene can open a gap in the spectrum of the order of $\sim20-50$meV
\cite{Kindermann,Justin,Jung,Pablo}, as recently observed in transport
measurements \cite{Hunt}. Due to the 1.8\% lattice mismatch between
graphene and its substrate \cite{Giovannetti,Yankowitz,Yang} and
possible twisted configurations between the two \cite{Giovannetti,Yankowitz,Yang,Sachs,Wallbank},
BN creates local potentials in graphene which modulate with the same
periodicity of the Moire pattern created by the two incommensurate
structures (Fig.1a) \cite{commensuration}. In the continuum limit,
the Hamiltonian of graphene in the presence of the BN substrate can
be generically written as 
\begin{equation}
\mathcal{H}=\int\mbox{d}^{2}r\sum_{\sigma}\sum_{\nu=\pm}\,\Psi_{\nu\sigma}^{\dagger}(\mathbf{r})[-vi\nabla\cdot\vec{\sigma}_{\nu}+\hat{A}_{\nu}(\mathbf{r})]\Psi_{\nu\sigma}(\mathbf{r})\,,\label{eq:Eq1}
\end{equation}
where $\Psi_{\nu}=(\psi_{\nu a},\psi_{\nu b})$ is a two component
spinor in the sublattice space in a given valley, $\vec{\sigma}_{\nu}=(\nu\sigma_{1},\sigma_{2})$
are the Pauli matrices defined for each valley ($\nu=\pm$), $v=6\mbox{eV}\AA$
is the Fermi velocity, $\sigma=\uparrow\downarrow$ indexes the spin
and  $\hat{A}_{\nu}(\mathbf{r})=\mu(\mathbf{r})\sigma_{0}+\nu\mathbf{A}(\mathbf{r})\cdot\vec{\sigma}_{\nu}+M(\mathbf{r})\sigma_{3}$\emph{
}are the local scalar, vector and mass term potentials induced by
the BN substrate, which spatially modulate with the Moire pattern.
In leading order, $\hat{A}_{\nu}(\mathbf{r})\approx\sum_{j=1}^{3}\cos(\mathbf{G}_{j}\cdot\mathbf{r})\,\hat{A}_{\nu}$,
where $\mathbf{G}_{j}$ are the reciprocal lattice vectors in the
Brillouin zone of the extended unit cell, and $\hat{A}_{\nu}$ parametrizes
the amplitudes of modulating potentials. 

As shown in previous tight binding models \cite{Kindermann,SM}, the
regions where the mass term changes sign forms a lattice of disconnected
quantum rings separating regions with opposite topological charges
\cite{Volovik}, as shown in Fig. 1b. In the presence of interactions,
the amplitude of the induced mass term is $M\equiv\mbox{max}[M(\mathbf{r})]\approx50-100$meV
\cite{SM,note4} for a Moire supercell with up to $140\mbox{\AA}$
in size \cite{Giovannetti,Yankowitz,Yang}. The real space topology
of those lines describes an insulating state in the bulk, unlike in
twisted graphene bilayers, where inversion symmetry is restored and
those gapless lines percolate into a metallic state with Dirac-like
quasiparticles \cite{Kindermann,Joao}. 

\begin{figure}[b]
\begin{centering}
\includegraphics[scale=0.29]{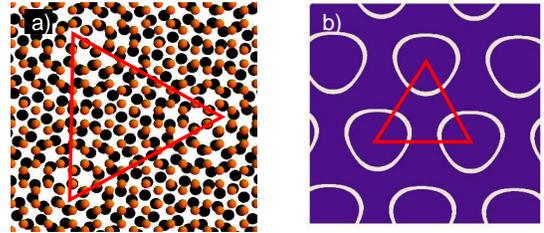}
\par\end{centering}

\caption{{\small{a) Moire pattern of graphene on top of boron nitride. b) Periodic
mass term potentials $M(\mathbf{r})$ induced on graphene by the BN
substrate. Solid rings: regions where the mass potential $M(\mathbf{r})$
crosses zero and changes sign \cite{SM}. }}}
\end{figure}

Those 1D circular domain walls can contain gapped low energy modes
when the amplitude of the induced mass term $M$ is larger than the
finite size gap $\approx v/(2\pi a)$ set by the radius of the rings
\cite{note}. In this regime, we find that Coulomb interactions lead
to spontaneous valley and spin polarization in those quantum rings,
which describe chiral loop currents in bulk. We develop an effective
lattice model and show that interactions lead to the subsequent formation
of macroscopic \emph{valley} and spin polarized low energy bands at
zero temperature. This exotic ordered state explicitly breaks TRS
and describes a ferromagnetic superlattice of spin and\emph{ valley}
local moments. We propose that the ferromagnetic valley order can
be detected with interferometry experiments and through the polar
Kerr effect, which measures the rotation of a linearly polarized beam
of light reflected on the sample. 

\emph{Toy model Hamiltonian.} In the presence of Coulomb interactions,
the mass term $M(\mathbf{r})$ is a relevant operator in the renormalization
group sense, while the scalar term $\mu(\mathbf{r})$ and the vector
potential term $\mathbf{A}(\mathbf{r})$ are not \cite{Foster}. The
latter are small compared to the mass term in the strong coupling
regime of the problem, which will be assumed \cite{Justin}. In this
regime, the mass term is the only relevant term and behaves as a periodic
function that changes sign in the nodal lines where $M(\mathbf{r})=0$. 

In cylindrical coordinates, $\mathbf{r}=(r,\theta),$ the mass term
profile for a single quantum ring can be approximated by a step function,
namely $M(r>a)=-M(r<a)=M$, where $a$ is the radius of the quantum
ring. The Hamiltonian matrix of a single ring can be written as $\hat{\mathcal{H}}(\mathbf{r})=\hat{\mathcal{H}}_{+}(\mathbf{r})\otimes\nu_{+}+\hat{\mathcal{H}}_{-}(\mathbf{r})\otimes\nu_{-}$,
where $\nu_{\pm}=(v_{0}\pm\nu_{3})/2$ are the valley projection operators,
with $\nu_{i}$ ($i=1,2,3$) as Pauli matrices, 
\begin{equation}
\hat{\mathcal{H}}_{+}(\mathbf{r})=\left(\begin{array}{cc}
M(r) & -i\mbox{e}^{-i\theta}(\partial_{r}-\frac{i}{r}\partial_{\theta})\\
-i\mbox{e}^{i\theta}(\partial_{r}+\frac{i}{r}\partial_{\theta}) & -M(r)
\end{array}\right),\label{eq:Ham2}
\end{equation}
is the Hamiltonian in valley $\nu=+$ and $\hat{\mathcal{H}}_{-}=\hat{\mathcal{H}}_{+}^{*}$
in the opposite valley (we set $v\to1$). The eigenvectors that satisfy
the equation $\hat{\mathcal{H}}(\mathbf{r})\Phi(\mathbf{r})=E\Phi(\mathbf{r})$
are the four component spinors $\Phi_{j,+}(\mathbf{r})=\left(\Psi_{j}(\mathbf{r}),\mathbf{0}\right)$
and $\Phi_{j,-}(\mathbf{r})=\left(\mathbf{0},\Psi_{j}^{*}(\mathbf{r})\right)$,
where
\begin{equation}
\Psi_{j}(\mathbf{r})=\left(\begin{array}{c}
F_{j}^{-}(r)\mbox{e}^{i(j-\frac{1}{2})\theta}\\
iF_{j}^{+}(r)\mbox{e}^{i(j+\frac{1}{2})\theta}
\end{array}\right),\label{eq:waveF}
\end{equation}
with $j=m+\frac{1}{2}$ the total angular momentum quantum number
($m\in\mathbb{Z}$), including orbital (valley) and pseudo-spin (sublattice)
degrees of freedom. Imposing the proper boundary conditions at $r=a$
and $r\to\infty$, $F_{j}^{\pm}(r)=A_{j}^{\pm}I_{|j\pm\frac{1}{2}|}(r\sqrt{M^{2}-E_{j}^{2}})\theta(a-r)+B_{j}^{\pm}K_{|j\pm\frac{1}{2}|}(r\sqrt{M^{2}-E_{j}^{2}})\theta(r-a)$,
with $I_{n}(x)$ and $K_{n}(x)$ as modified Bessel functions, and
$A_{j}^{\pm},$ $B_{j}^{\pm}$ the proper coefficients (see Fig. 2a).
For $Ma\gg1$ the wave functions are sharply peaked at $r=a$, and
the states are localized at the domain wall where the mass term changes
sign. In the opposite regime, when $Ma$ is of the order 1, the electrons
can tunnel across the center of the ring and their wavefunctions become
extended over the area of each ring, as in a quantum dot. In any case,
the energy spectrum of the $j$ energy level is set by the condition
\begin{equation}
\frac{1}{4M^{2}}\prod_{s=\pm1}\partial_{a}\ln\frac{K_{|j+\frac{s}{2}|}(\sqrt{M^{2}-E_{j}^{2}}a)}{I_{|j+\frac{s}{2}|}(\sqrt{M^{2}-E_{j}^{2}}a)}=1,\label{eq:constrraint}
\end{equation}
which gives a discrete spectrum of gapped low energy modes confined
inside the quantum rings, as shown in Fig. 2b as a function of $Ma$. 

The energy spectrum inside the gap is particle hole symmetric, with
$j=m+\frac{1}{2}>0$ describing positive energy states and $j<0$
describing negative energy ones. The red curves correspond to $|j|=\frac{1}{2}$
states, while the other three curves describe $|j|=\frac{3}{2},\,\frac{5}{2}$
and $\frac{7}{2}$ states respectively, the outer curves having higher
$|j|$. In all cases, there is a critical value of $Ma$ below which
a given mode dives in the continuum of the band outside the gap. Inside
the gap, those discrete levels are sharply defined and describe the
circular motion of electrons physically confined inside the quantum
rings shown in Fig. 1b. All levels have four-fold degeneracy, with
two spins and two valleys. Their spin and orbital degeneracies can
be lifted by repulsive interactions, which can give rise to locally
polarized states. 

\emph{Valley and spin polarized states. }The Coulomb interaction between
the electrons is 
\begin{equation}
\mathcal{H}_{C}=\frac{1}{2}\int\mbox{d}^{2}r\mbox{d}^{2}r^{\prime}\,\hat{\rho}(\mathbf{r})V(\mathbf{r}-\mathbf{r}^{\prime})\hat{\rho}(\mathbf{r}^{\prime}),\label{Hc}
\end{equation}
where $V(\mathbf{r}-\mathbf{r}^{\prime})=e^{2}/(\kappa|\mathbf{r}-\mathbf{r}^{\prime}|)$,
with $e$ the electric charge, $\kappa\approx2.5$ the dielectric
constant due to the BN substrate and $\hat{\rho}(\mathbf{r})=\sum_{\sigma}\Theta_{\sigma}^{\dagger}(\mathbf{r})\Theta_{\sigma}(\mathbf{r})$
is a density operator defined in terms of the field operators $\Theta_{\sigma}(\mathbf{r})\equiv\sum_{\nu,j}\Phi_{\nu,j}(\mathbf{r})c_{\nu,\sigma,j},$
where $c_{\nu,\sigma,j}$ describes an annihilation operator with
spin $\sigma$ on a given valley and angular momentum state $j=m+\frac{1}{2}$. 

\begin{figure}
\begin{centering}
\includegraphics[scale=0.31]{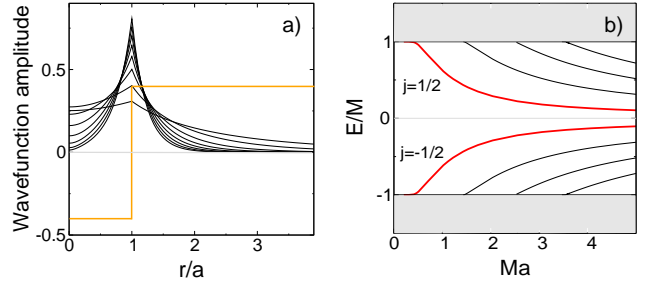}
\par\end{centering}

\caption{{\small{a) Amplitude of the wavefunctions around the domain wall set
by the the quantum rings for $Ma$ ranging from $0.7$ to 3.5 for
$|j|=\frac{1}{2}$. Orange line: profile of mass term potential $M(r)$
for a single quantum ring within the toy model. b) Gapped low energy
modes $j=m+\frac{1}{2},$ for $m=-4,\,-3\ldots3$ from bottom to the
top curves. All modes are four-fold degenerate. The red lines indicate
the $j=\pm\frac{1}{2}$ states.}}}

\end{figure}

The Coulomb interaction at the $j$-th level in a given quantum ring
can be written as 
\begin{equation}
\mathcal{H}_{U}=U\hat{n}_{\uparrow}\hat{n}_{\downarrow}+U\sum_{\sigma}\hat{n}_{+,\sigma}\hat{n}_{-,\sigma},\label{Hc1}
\end{equation}
where 
\begin{equation}
U=\int\mbox{d}^{2}r\mbox{d}^{2}r^{\prime}|\Phi_{\nu,j}(\mathbf{r})|^{2}V(\mathbf{r}-\mathbf{r}^{\prime})|\Phi_{\nu^{\prime},j}(\mathbf{r}^{\prime})|^{2}\label{eq:U}
\end{equation}
is the valley independent Hubbard coupling and $\hat{n}_{\sigma}=\sum_{\nu}\hat{n}_{\nu,\sigma}$
describes the occupation of the $j$-th state in terms of $c$ operators
($j$ level indexes omitted). The Hubbard $U$ term is shown in Fig.
3a as a function of $Ma$ and shows a non-monotonic behavior, reflecting
the crossover of the wavefunctions for $Ma\lesssim1$, when the electrons
can easily tunnel through the center of the quantum rings. At $Ma\lesssim0.4$,
the $|j|=\frac{1}{2}$ states merge the continuum, and the toy model
description breaks down. The exchange interaction in a given ring
is identically \emph{zero} due to the orthogonality of the eigenspinors
in different valleys, $\Phi_{+}^{\dagger}(\mathbf{r})\Phi_{-}(\mathbf{r})=0$
\cite{note1}. The problem of an isolated quantum ring in a given
$j$ state is dual to the problem of a \emph{doubly} degenerate orbital
with spin $\frac{1}{2}$, and can be mapped in the Coqblin-Bladin
model for two degenerate orbitals \cite{Coqblin}. 

\begin{figure}[t]
\begin{centering}
\includegraphics[scale=0.32]{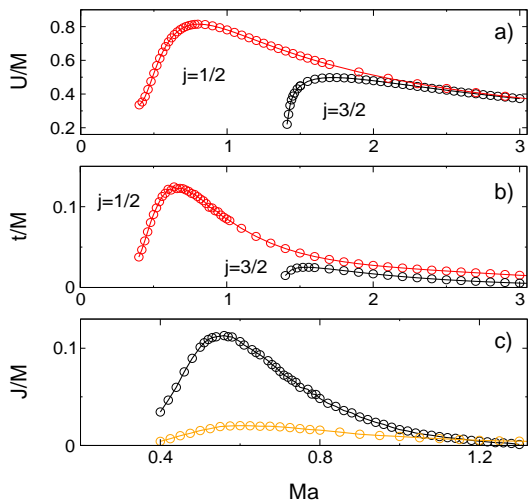}
\par\end{centering}

\caption{{\small{a) Scaling of the Hubbard energy $U$ and b) the hopping energy
between nearest ring $t$ vs $Ma$. Red curves: $|j|=\frac{1}{2}$
levels; black curves: $|j|=\frac{3}{2}$. c) Exchange coupling $J$
(black circles) and superexchange coupling $t^{2}/U$ (orange) in
$M$ units vs. $Ma$ in the $|j|=\frac{1}{2}$ state. For $0.4\lesssim Ma\lesssim1$.1,
the system shows ferromagnetic valley order (see text). }}}
\end{figure}

At the mean field level, the effective Hamiltonian of the $j$-th
state with bare energy $E_{0}$ is $H_{L}=\sum_{\nu\sigma}E_{\nu\sigma}\hat{n}_{\nu,\sigma},$
where 
\[
E_{\nu,\sigma}=E_{0}+U\sum_{\nu^{\prime}}n_{\nu^{\prime},-\sigma}+Un_{-\nu,\sigma}
\]
 is the renormalized energy due to interactions. The occupation of
the four degenerate states $n_{\nu,\sigma}$ $(\nu=\pm,\sigma=\uparrow\downarrow)$
in the $j$-th level can be calculated self-consistently from the
Greens function of the localized $c$ electrons, $G_{\sigma,\nu}(\omega)=(\omega-E_{\sigma,\nu}+i\delta)^{-1},$
namely $n_{\nu,\sigma}=\langle\hat{n}_{\nu,\sigma}\rangle=-\frac{1}{\pi}\mbox{Im}\int_{-\infty}^{\mu}\mbox{d}\omega\, G_{\sigma,\nu}(\omega)$,
with $\mu$ the chemical potential. When the repulsion $U$ is the
dominant energy scale, the lowest energy solution is a state where
$n_{\nu,\sigma}=N_{+}$ and $n_{\nu,-\sigma}=n_{-\nu,\sigma}=n_{-\nu,-\sigma}=N_{-}$,
which is spin and valley polarized for $N_{+}\neq N_{-}$ \cite{Coqblin}.
In this regime, 
\begin{equation}
N_{s}=\frac{1}{2}-\frac{1}{\pi}\mbox{arctan}\left(\frac{2N_{-}+N_{-s}-x}{y}\right),\label{eq:Ns}
\end{equation}
with $s=\pm$, where $x=(\mu-E_{0})/U$ and $y=\delta/U$, with $\delta$
the level broadening. In the limit $y\to0$, when the levels are sharply
defined inside the gap, and $E_{0}<\mu<E_{0}+U$, the lowest energy
solution is a maximally spin and valley polarized state with $N_{+}=1$
and $N_{-}=0$. This state describes a lattice of isolated quantum
rings with \emph{random} spin polarized circulating charge currents. 

\emph{Nearly flat bands.} The effective tight binding Hamiltonian
for the $c$ electrons moving in a triangular superlattice of quantum
rings is $\mathcal{H}_{eff}=\mathcal{H}_{t}+\sum_{i}\mathcal{H}_{U,i}+\sum_{\langle i,j\rangle}\mathcal{H}_{C,ij}$,
where
\begin{equation}
\mathcal{H}_{t}=t\sum_{\langle ij\rangle}\sum_{\nu\sigma}c_{i,\nu,\sigma}^{\dagger}c_{j,\nu,\sigma}\label{eq:Ht2}
\end{equation}
is the kinetic energy of the electrons, with $c_{i}$ the annihilation
operator for an electron in a quantum ring centered at $\mathbf{R}_{i}$,
and $\langle ij\rangle$ indexes nearest neighbor (NN) sites. $t$
is the hopping energy betwen NN rings, $t_{ij}=\int\mbox{d}^{2}r\Phi_{\nu}^{\dagger}(\mathbf{r}_{i})\delta\hat{M}(\mathbf{r})\Phi_{\nu}(\mathbf{r}_{j})$,
with $\mathbf{r}_{i}\equiv\mathbf{r}-\mathbf{R}_{i}$, where $\delta\hat{M}(\mathbf{r})=\delta M(\mathbf{r})\sigma_{3}\otimes\nu_{0}$
is the the mass potential that restores the periodicity of the superlattice
when added to the step function potential $M(\mathbf{r})=M\,\mbox{sign}(r-a)$
due to one isolated quantum ring at the origin. The second term, $\mathcal{H}_{U,i}$,
is the on-site Coulomb interaction (\ref{Hc1}) on a given site $i$
in the superlattice, and is defined by $\hat{n}_{i,\nu,\sigma}$ density
operators. The third one, $\mathcal{H}_{C,ij}$, describes the Coulomb
interaction (\ref{Hc}) between different superlattice sites. 

The hopping amplitude $t$ shown in Fig. 3b has a non-monotonic behavior
as a function of $Ma$ which mimics the behavior of the Hubbard $U$
coupling, and is typically one order of magnitude smaller than the
Coulomb interaction, $U/|t|\gtrsim7$. In particular, for $M\approx50-100$meV
\cite{note4} and for a typical superlattice size of $3a\approx140\mbox{\AA}$
\cite{Yankowitz,Yang} in graphene nearly aligned with BN, $Ma\in[0.4,\,0.8]$,
which corresponds to a ratio $7\lesssim U/t\lesssim9$. At quarter
filling ($\mu=0$), that suggests that correlations keep the gapped
1D modes inside the rings strongly localized. In order to account
for the macroscopic order of the chiral loop currents in bulk, we
examine the electronic correlations among the rings. 

As electrons hop between different superlattice sites, the on-site
correlation tends to align either their valley or spin quantum numbers
antiferromagnetically due to Pauli principle, in order to reduce the
energy cost of the kinetic energy. In second order of perturbation
theory, the super-exchange interaction among the rings is given by
$\mathcal{H}_{S}=\mathcal{H}_{t}\mathcal{H}_{U}^{-1}\mathcal{H}_{t}+\mathcal{O}(t^{4})$,
or equivalently $\mathcal{H}_{s}=-(t{}^{2}/U)\sum_{\langle ij\rangle}\sum_{\{\nu\}\{\sigma\}}c_{i,\nu,\sigma}^{\dagger}c_{j,\nu,\sigma}c_{j,\nu^{\prime},\sigma^{\prime}}^{\dagger}c_{i,\nu^{\prime},\sigma^{\prime}}$
\cite{Kugel}. This term maps into the SU(4) Heisenberg Hamiltonian
\begin{equation}
\mathcal{H}_{s}=4\frac{t{}^{2}}{U}\sum_{\langle ij\rangle}\left(\frac{1}{4}+\boldsymbol{\tau}_{i}\cdot\boldsymbol{\tau}_{j}\right)\left(\frac{1}{4}+\mathbf{S}_{i}\cdot\mathbf{S}_{j}\right)\label{eq:se-1}
\end{equation}
in a triangular lattice, where $\mathbf{S}_{i}$ is a spin $\frac{1}{2}$
operator on site $i$ and $\boldsymbol{\tau}_{i}$ the equivalent
pseudo-spin operator, which acts in the valleys. This Hamiltonian
is frustrated and is expected to describe a \emph{spin-orbital liquid}
in the ground state \cite{Penc}.

The Coulomb interaction between rings, $\mathcal{H}_{C,ij}$, follows
directly from Hamiltonian (\ref{Hc}) by properly including the superlattice
into the definition of the field operators $\Theta_{\sigma}(\mathbf{r})=\sum_{\nu,i}\Phi_{\nu}(\mathbf{r}_{i})c_{i,\nu\sigma}$.
This term can be written explicitly in the form of the exchange interaction
$\mathcal{H}_{e}=J\sum_{\langle ij\rangle}\sum_{\{\nu\}\{\sigma\}}c_{i,\nu,\sigma}^{\dagger}c_{j,\nu^{\prime},\sigma^{\prime}}^{\dagger}c_{i\nu^{\prime},\sigma^{\prime}}c_{j,\nu,\sigma},$
where $J>0$ is the exchange coupling, $J_{ij}\!=\!\frac{1}{2}\int\mbox{d}^{2}r\mbox{d}^{2}r^{\prime}\,\Phi_{\nu}^{\dagger}(\mathbf{r}_{i})\Phi_{\nu}(\mathbf{r}_{j})V(|\mathbf{r}-\mathbf{r}^{\prime}|)\Phi_{\nu^{\prime}}^{\dagger}(\mathbf{r}_{j}^{\prime})\Phi_{\nu^{\prime}}(\mathbf{r}_{i}^{\prime}),$
and can also be cast into the form of an SU(4) Heisenberg model
\begin{equation}
\mathcal{H}_{e}=-4J\sum_{\langle ij\rangle}\left(\frac{1}{4}+\boldsymbol{\tau}_{i}\cdot\boldsymbol{\tau}_{j}\right)\left(\frac{1}{4}+\mathbf{S}_{i}\cdot\mathbf{S}_{j}\right).\label{eq:exchange}
\end{equation}

When $J>t{}^{2}/U$, the exchange coupling dominates and drives the
system into a \emph{spin-valley} \emph{ferromagnetic} state with true
long range order at zero temperature, giving rise to \emph{spin-valley}
polarized low energy bands. At strong enough coupling, those bands
are expected to become \emph{nearly flat}. In the corresponding midgap
band formed by $j=-\frac{1}{2}$ levels, the spin-valley ferromagnetic
state emerges for $0.4\lesssim Ma\lesssim1.1$, as shown in Fig. 3c.
In this interval, $J\lesssim0.1M\sim5-10$meV. Although knowing the
exact polarization of the low energy bands requires self-consistently
solving a non-trivial strongly correlated problem, when $U\gg t$
interactions are strong and lead to a net spin-valley polarization
in the midgap states at zero temperature. 

\emph{Experimental observation. }In the valley ferromagnetic state,
the loop currents in bulk break TRS and produce a ferromagnetic lattice
of local magnetic moments $\approx\mu_{B}$, with $\mu_{B}$ a Bohr
magneton. An external magnetic field $\mathbf{H}$ couples with the
spin-valley moments through the Zeeman coupling, $\mathcal{H}_{Z}=-2\mu_{B}(\boldsymbol{\tau}+\mathbf{S})\cdot\mathbf{H}$.
Due to the proximity of the ordered ground state at $T=0$, a very
weak applied magnetic field $\mu_{B}H_{z}\sim0.01k_{B}T$ can produce
a large spin-valley magnetization $\sim1\mu_{B}$ \cite{Antsygina}.
For instance, at temperatures $T\sim0.01J/k_{B}\lesssim1$K, the required
applied field can be smaller than $H_{z}\lesssim0.01$ T. In this
regime, this state can generate a macroscopic flux $\Phi$ that is
proportional to the spin-valley polarization. This flux can be detected
with standard superconducting quantum interference devices placed
on top of graphene \cite{Sepioni}, as illustrated in Fig. 4a.

\begin{figure}[t]
\begin{centering}
\includegraphics[scale=0.31]{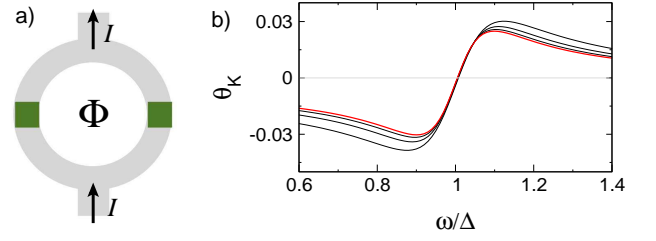}
\par\end{centering}

\caption{{\small{a) Magnetic flux $\Phi$ produced by the valley ferromagnetic
state, measured by an interference device (gray region) on top of
graphene. The supercurrent $I$ splits between two Josephson junctions
(on green). b) Polar Kerr angle $\theta_{K}$ in radians versus photon
energy $\omega$ normalized by the optical gap $\Delta$ for transitions
between $j=\pm\frac{1}{2}$ energy flat bands. Curves for $Ma=0.5,\,0.6,\,0.7$
and $0.8$ (red) (see text). }}}
\end{figure}

When linearly polarized light is applied over an atomically thin medium
that breaks TRS, the light polarization rotates by the Kerr angle
$\theta_{K}(\omega)=8\pi/[c(n^{2}-1)]\mbox{Re}\,\sigma_{xy}(\omega)$
\cite{Nandkishore}, where $\sigma_{xy}$ is the anomalous Hall conductivity
\cite{Nagaosa}, which is proportional to the \emph{valley} polarization
\cite{note3}, $c$ is the speed of light and $n\approx2.5$ is the
refraction index of the BN substrate. Within the toy model (\ref{eq:Ham2}),
the anomalous Hall conductivity can be derived by defining the electronic
Green's function $G_{\nu}(\mathbf{r},\mathbf{r}^{\prime},\omega)=\sum_{j,\mathbf{k}}\Phi_{\nu,j,\mathbf{k}}(\mathbf{r})\Phi_{\nu,j,\mathbf{k}}^{\dagger}(\mathbf{r}^{\prime})/(\omega-E_{j}+i\gamma)$
in terms of the Bloch waves in the superlattice for a given valley
$\nu$, $\Phi_{\nu,j,\mathbf{k}}(\mathbf{r})=\sum_{i}\Phi_{\nu,j}(\mathbf{r}_{i})\mbox{e}^{i\mathbf{k}\cdot\mathbf{R}_{i}}$.
For simplicity, we assume that $E_{j}$ is the energy of a dispersionless
flat band indexed by the angular momentum state $j$ and $\gamma$
is the inverse of the quasiparticle lifetime. 

The anomalous Hall conductivity in valley $\nu=+$ follows from the
current-current correlation function $\Pi_{xy}(\mathbf{r},\mathbf{r}^{\prime},\omega)=e^{2}\mbox{tr}\!\int V_{+,x}G_{+}(\mathbf{r},\mathbf{r}^{\prime},\omega^{\prime})V_{+,y}G_{+}(\mathbf{r}^{\prime},\mathbf{r},\omega^{\prime}\!+\omega)\,\mbox{d}\omega^{\prime}/2\pi$,
with $V_{\nu,i}=v(\sigma_{\nu,i}\otimes\nu_{0})$ \cite{Nagaosa}.
In momentum space, the optical Hall conductivity is $\sigma_{xy}(\omega)=(i/\omega)\lim_{\mathbf{q}\to0}\Pi_{xy}(\mathbf{q},-\mathbf{q},\omega)$.
The transitions between the valley polarized $j=\pm\frac{1}{2}$ bands
dominate the Hall response for frequencies near the optical gap $\Delta=2E_{j=\frac{1}{2}}$.
In this frequency range ($\sim10^{13}$Hz), the zero temperature response
is \cite{SM} 
\begin{equation}
\sigma_{xy}(\omega)\approx c_{0}^{2}\frac{e^{2}}{h}\frac{(\hbar v\Lambda)^{2}}{(\hbar\omega+i\gamma)^{2}-\Delta^{2}}\label{eq:sigma}
\end{equation}
restoring $\hbar$, where $\Lambda\sim2\pi/(3a)$ is the size of the
Moire Brillouin zone and $c_{0}=\int\mbox{d}^{2}r\, F_{\frac{1}{2}}^{-}(r)F_{-\frac{1}{2}}^{+}(r)\approx0.81$. 

For $\gamma\sim15$meV \cite{Zhou} and $\hbar v\Lambda\approx0.26$eV,
which corresponds to a Moire unit cell of $140\mbox{\AA}$, the Kerr
angle is $\theta_{K}\sim10^{-2}$ radians for maximal valley polarization,
as shown in Fig. 4b. For a weak valley magnetization of $0.1\mu_{B}$,
the Kerr rotation is $\theta_{K}\sim10^{-3}$, which is still very
large. This effect that can be detected with THz/infrared Kerr experimental
setups \cite{Zhou}. In the visible range, Hall Kerr measurements
are extremely sensitive and are able to detect rotations as small
as $\theta_{K}\sim10^{-9}$ radians \cite{Kapitulnik}. By changing
the occupation of the midgap states, the valley ferromagnetic order
can be controlled with a gate voltage. This exotic state has clear
experimental signatures and can lead to the experimental realization
of valley order in graphene at low temperature and weak applied magnetic
fields \cite{Amet}. 

\emph{Acknowledgements. }We thank F. Guinea, E. Andrei, I. Martin,
F. Mila, T. G. Rappoport, A. Del Maestro, K. Mullen, and A. Sandvik
for discussions. BU acknowledges University of Oklahoma and NSF Career
grant DMR-1352604 for support. VNK was supported by US DOE grant DE-FG02-08ER46512,
and MK by NSF grant DMR-1055799.


\onecolumngrid
\newpage

\begin{center}
\textbf{\large{Suplementary Materials for ``Valley order and loop
currents in graphene on hexagonal boron nitride''}}{\large{}}\\

\par\end{center}{\large \par}

\begin{center}
Bruno Uchoa, Valeri  N. Kotov and M. Kindermann
\par\end{center}


\section{Effective Hamiltonian in the Continuum}

In the absence of interactions, the Hamiltonian of a two layer system
is described by three terms,
\begin{equation}
\mathcal{H}=\mathcal{H}_{1}+\mathcal{H}_{2}+\mathcal{H}_{1-2}.\label{eq:a}
\end{equation}
 The first two terms describe the kinetic energy in each of the layers
in separate, which in tight binding form is 
\[
\mathcal{H}_{l}=-t\sum_{\langle ij\rangle}\bar{\psi}_{l,a}^{\dagger}(\mathbf{R}_{i})\bar{\psi}_{l,b}(\mathbf{R}_{j})+(\mu_{l}+m_{l})\sum_{i}n_{l}(\mathbf{R}_{i})
\]
with $l=1,2$ indexing the different layers, where $\bar{\psi}_{a}$
($\bar{\psi}_{b})$ is an annihilation operator acting on sublattice
$A$ ($B$) of each layer, $n_{l}$ is an on-site density operator
on layer $l$, $t\sim3$eV is the in-plane hopping energy, $\mu_{l}$
is the chemical potential, $m_{l}$ is the intrinsic mass gap of each
layer and $\langle ij\rangle$ indicate sum over the nearest neighbor
sites. Spin indexes will be omitted. For graphene on BN, we have $\mu_{1}=m_{1}=0$,
$\mu_{2}\equiv V$ and $m_{2}\equiv m\neq0$.

The $\bar{\psi}_{a,b}(\mathbf{r})$ operators can be written in a
basis of Bloch wave functions as \cite{kindermann2,Mele2} 
\begin{equation}
\bar{\psi}_{x,l}(\mathbf{R})=\sum_{\nu=\pm}\phi_{x,l,\nu}(\mathbf{R})\,\psi_{x,l,\alpha}(\mathbf{R}),\label{eq:psi}
\end{equation}
where $\psi_{x}$ $(x=A,B)$ define the new fermionic operators, 
\[
\phi_{x,\nu,l}(\mathbf{R})=\frac{1}{\sqrt{3}}\sum_{a=1}^{3}\mbox{e}^{i\mathbf{K}_{\nu,l}^{a}\cdot(\mathbf{R}-\mathbf{R}_{x,l}^{0})}
\]
 is the corresponding Block wave function for each layer, $\mathbf{R}_{x,l}^{0}$
gives the position of a given arbitrary site on sublattice $x$ and
layer $l$, and $\nu=\pm$ correspond to the two different valleys,
each one represented by three distinct $\mathbf{K}_{\nu,l}^{a}$ vectors
located at the corners of the Brillouin zone . In the continuum limit,
\begin{equation}
\mathcal{H}_{l}=\!\sum_{\nu=\pm}\int\!\!\mbox{d}^{2}r\,\Psi_{l,\nu}^{\dagger}(\mathbf{r})\!\left[-iv\vec{\sigma}_{\nu}\cdot\nabla+V_{l}\sigma_{0}+m_{l}\sigma_{3}\right]\!\Psi_{l,\nu}(\mathbf{r}),\label{eq:Ho}
\end{equation}
where $\Psi=(\psi_{a},\psi_{b})$ is a two component spinor in the
sublattice space of each layer, $\vec{\sigma}_{\nu}=(\nu\sigma_{1},\sigma_{2})$
are the Pauli matrices defined for each valley and $v=6\mbox{eV}\AA$
is the Fermi velocity, and $V_{l}$ are the local scalar potential
in both layers. 

The third term in (\ref{eq:a}), $H_{1-2}$, describes the electronic
hopping between the two layers, which in the continuum limit is described
by 
\begin{equation}
\mathcal{H}_{1-2}=\int\mbox{d}^{2}r\,\sum_{\nu=\pm}\Psi_{1,\nu}^{\dagger}(\mathbf{r})\hat{t}_{\nu,\perp}(\mathbf{r})\Psi_{2,\nu}(\mathbf{r})+h.c,\label{eq:122}
\end{equation}
where 
\[
t_{\nu,\perp}^{x,y}(\mathbf{r})=t_{\perp}\phi_{x,1,\nu}^{*}(\mathbf{r})\phi_{y,2,\nu}(\mathbf{r})
\]
 is the \emph{interlayer} hopping matrix, with $t_{\perp}\sim0.4$
eV the hopping amplitude \cite{Giovanetti2}. 

The effective Hamiltonian of the gapless layer $1$ (graphene) can
be computed directly by integrating out the electrons in the second
layer, $\bar{\mathcal{H}}_{1}=\mathcal{H}_{1}+\delta\mathcal{H}_{1}$
where the second term describes the effective local potentials induced
by layer 2. In lowest order in perturbation theory \cite{Kindermann3}, 

\begin{equation}
\delta\mathcal{H}_{1}=\int\mbox{d}\mathbf{r}\sum_{\nu=\pm}\Psi_{1,\nu}^{\dagger}(\mathbf{r})\hat{t}_{\nu,\perp}(\mathbf{r})\hat{M}\hat{t}_{\nu,\perp}^{\dagger}(\mathbf{r})\Psi_{1,\nu}(\mathbf{r}),\label{eq:H_perp-1}
\end{equation}
where 
\begin{equation}
\hat{M}=\frac{1}{\omega-V+m}\left(\begin{array}{cc}
\eta & 0\\
0 & 1
\end{array}\right)\label{eq:M}
\end{equation}
where $\omega$ is the interlayer applied bias voltage, and 
\[
\eta\equiv-\frac{m-V-\omega}{m+V+\omega}\approx-\frac{1.5-\omega}{3.1+\omega}.
\]

In the first star approximation\cite{Mele2}, where backscattering
process are restricted to the first BZ of the extended unit cell,
the spacial modulation of those fields can be approximated to a sum
over the three reciprocal lattice vectors $\mathbf{G}_{j}$ of the
extended unit cell \cite{kindermann2}, 
\begin{equation}
\hat{A}_{\nu}(\mathbf{r})=\hat{t}_{\nu,\perp}(\mathbf{r})\hat{M}\hat{t}_{\nu,\perp}^{\dagger}(\mathbf{r})\approx\sum_{j=1}^{3}\cos(\mathbf{G}_{j}\cdot\mathbf{r})\hat{A}_{\nu},\label{eq:tMt}
\end{equation}
where $\hat{A}_{\nu}$ is in the form $\hat{A}_{\nu}\equiv\mu\sigma_{0}+\mathbf{A}\cdot\vec{\sigma}_{\nu}+M\sigma_{3}$.

For graphene at half filling on BN, the microscopic parameters can
be extracted from ab initio calculations. The intrinsic BN gap is
$m\approx2.3$ eV and $V\approx0.8$eV \cite{Giovanetti2}. At zero
interlayer bias, $\eta=-0.5$,  which describes Fig. 1 of the main
text at small twist angles. 

For $t_{\perp}\approx0.4$eV and zero bias, the maximal allowed amplitude
for the mass term is $M\approx t_{\perp}^{2}/(m-V)\sim100$meV. In
the absence of interactions, one may adopt a conservative estimate
of $M\approx50$ meV, which is consistent with recent ab initio results
\cite{Sachs-1}. Many-body effects can significantly renormalize $M$
and make it substantially larger \cite{KotovSM,Justin-1}. In the
manuscript, we consider the effects of renormalized low energy bands
corresponding to an amplitude of the mass term in the range $M\sim50-100$meV.

\section{Anomalous Hall Conductivity}

The Bloch wave functions for electrons in a lattice of quantum rings
is
\begin{equation}
\Psi_{j,\mathbf{k}}(\mathbf{r})=\sum_{\mathbf{R}}\Psi_{j}(\mathbf{r}-\mathbf{R})\mbox{e}^{i\mathbf{k}\cdot\mathbf{R}}\label{eq:Psi}
\end{equation}
where $\mathbf{R}$ indexes the superlattice sites, and 
\begin{equation}
\Psi_{j}(\mathbf{r})=\left(\begin{array}{c}
F_{j}^{-}(r)\mbox{e}^{i(j-\frac{1}{2})\theta}\\
iF_{j}^{+}(r)\mbox{e}^{i(j+\frac{1}{2})\theta}
\end{array}\right),\label{eq:waveF}
\end{equation}
is the wavefunction in a given ring on valley $v=+$. The real space
Green's function is 
\begin{eqnarray*}
G(\mathbf{r},\mathbf{r}^{\prime},i\omega) & = & \sum_{j,\mathbf{k}}\frac{\Psi_{j,\mathbf{k}}(\mathbf{r})\Psi_{j,\mathbf{k}}^{\dagger}(\mathbf{r}^{\prime})}{i\omega-\epsilon_{j}}\\
 & = & \sum_{j}\sum_{\mathbf{R}}\frac{\Psi_{j}(\mathbf{r}-\mathbf{R})\Psi_{j}^{\dagger}(\mathbf{r}^{\prime}-\mathbf{R})}{i\omega-\epsilon_{j}},
\end{eqnarray*}
with $\epsilon_{j}$ the energy of a dispersionless flat band $j$.
The Fourier transform of the Green's function in momentum space is
\[
G(\mathbf{p},\mathbf{p}^{\prime},i\omega)=\delta_{\mathbf{p},\mathbf{p}^{\prime}}\sum_{j}\int\!\mbox{d}\mathbf{r}\,\mbox{d}\mathbf{r}^{\prime}\frac{\Psi_{j}(\mathbf{r})\Psi_{j}^{\dagger}(\mathbf{r}^{\prime})}{i\omega-\epsilon_{j}}\,\mbox{e}^{i\mathbf{p}\cdot(\mathbf{r}-\mathbf{r}^{\prime})}.
\]
The current-current correlation function is
\begin{equation}
\Pi_{xy}(\mathbf{q},\mathbf{q}^{\prime},i\omega)=\frac{e^{2}v^{2}}{\beta}\sum_{i\omega^{\prime}}\sum_{\mathbf{p},\mathbf{p}^{\prime}}\mbox{tr}\left[\sigma_{x}G(\mathbf{p}^{\prime}+\mathbf{q},\mathbf{p}-\mathbf{q}^{\prime},i\omega^{\prime}+i\omega)\sigma_{y}G(\mathbf{p},\mathbf{p}^{\prime},i\omega^{\prime})\right],\label{Pi}
\end{equation}
with $\beta$ the inverse of temperature. Accounting only for transitions
between the $j=\pm\frac{1}{2}$ states, which are dominant at frequencies
near the optical gap $\omega\sim2\epsilon_{j=\frac{1}{2}},$ 
\begin{equation}
\Pi_{xy}(\mathbf{q},\mathbf{q}^{\prime},i\omega)\approx\frac{i}{4\pi}e^{2}(v\Lambda)^{2}c_{0}^{2}\,\delta_{\mathbf{q},-\mathbf{q}^{\prime}}\left[\frac{1}{i\omega-\epsilon_{\frac{1}{2}}+\epsilon_{-\frac{1}{2}}}+\frac{1}{i\omega+\epsilon_{\frac{1}{2}}-\epsilon_{-\frac{1}{2}}}\right]\left[n_{F}(\epsilon_{\frac{1}{2}})-n_{F}(\epsilon_{-\frac{1}{2}})\right]\mbox{e}^{i\mathbf{q}\cdot(\mathbf{r}-\mathbf{r}^{\prime})},\label{Pixy3}
\end{equation}
where $\Lambda\approx2\pi/(3a)$ is a momentum cut-off set by the
size of the Moire BZ, $c_{0}\equiv\int_{0}^{\infty}\mbox{d}^{2}r\, F_{\frac{1}{2}}^{-}(r)F_{-\frac{1}{2}}^{+}(r)\sim0.81$,
and $n_{F}$ is the Fermi distribution. The optical Hall conductivity
follows from $\sigma_{xy}(\omega)=(i/\omega)\lim_{\mathbf{q}\to0}\Pi_{xy}(\mathbf{q},-\mathbf{q},\omega+i0^{+})$.


\begin{thebibliography}{10}
\bibitem{note4}In the non-interacting picture, $M_{0}\sim$ 50meV
for large Moire unit cells \cite{SM}. RG results indicate that $M=M_{0}(\lambda)^{\beta}$,
with $\beta=16/(\pi^{2}N)\sim0.4$ \cite{Justin,Kotov2}, and $1<\lambda\lesssim3a/a_{0}\approx100$
sets the length scale of the RG flow, which stops at the size of the
Moire unit cell, with $a_{0}\sim1.42\mbox{\AA}$ the lattice parameter.
Hence, $M/M_{0}\approx1-6$. In experiment, the renormalization is
limited by infrared cut-offs set by disorder and screening from metallic
contacts.
\bibitem{note1} In Eq. (\ref{Hc}), the exchange contribution for
two electrons in the same ring is $J\sum_{\sigma,\sigma^{\prime}}c_{+,\sigma}^{\dagger}c_{-,\sigma^{\prime}}^{\dagger}c_{+,\sigma^{\prime}}c_{-,\sigma}$,
where $J=\frac{1}{2}\int\mbox{d}^{2}r\mbox{d}^{2}r^{\prime}\,\Phi_{+}^{\dagger}(\mathbf{r})\Phi_{-}(\mathbf{r})V(\mathbf{r}-\mathbf{r}^{\prime})\Phi_{-}^{\dagger}(\mathbf{r}^{\prime})\Phi_{+}(\mathbf{r}^{\prime})\equiv0$.
\bibitem{note3}Since the spin-orbit coupling in graphene is tipically
small, the valley order should dominate the magneto-optical response.
\end{thebibliography}
\end{document}